\documentclass[12pt,preprint]{aastex}
\newcommand{\be}{\begin{equation}}
\newcommand{\ee}{\end{equation}}
\newcommand{\bea}{\begin{eqnarray}}
\newcommand{\eea}{\end{eqnarray}}

\begin{document}

\title{Dust Motions Driven by MHD Turbulence}

\author{A. Lazarian \& Huirong Yan}

\affil{Department of Astronomy, University of Wisconsin, 475 N. Charter
St., Madison, WI53706; lazarian, yan@astro.wisc.edu}

\begin{abstract}
We discuss the relative grain motions due to MHD turbulence in interstellar
medium. It has been known for decades that turbulent drag is an efficient
way to induce grain relative motions. However, earlier treatments
disregarded magnetic field and used Kolmogorov turbulence. Unlike
hydro turbulence, MHD turbulence is anisotropic on small scales. Moreover,
compressible modes are important for MHD and magnetic perturbations
can directly interact with grains. We provide calculations of grain
relative motion for realistic interstellar turbulence driving that
is consistent with the velocity dispersions observed in diffuse gas
and for realistic grain charging. We account for the turbulence cutoff
arising from abmipolar drag. Our results on grain shattering are consistent
with the customary accepted cutoff size. We obtain grain velocities
for turbulence with parameters consistent with those in HI and dark
clouds. These velocities are smaller than those in earlier papers,
where MHD effects were disregarded. Finally, we consider grain velocities
arising from photoelectric emission, radiation pressure and H\( _{2} \)
thrust. These are lower than relative velocities induced by turbulence.
We conclude that turbulence should prevent these mechanisms from segregating
grains by size. 
\end{abstract}


\section{Introduction}

Dust is an important constituent of the interstellar medium (ISM).
It interferes with observations in the optical range, but provides
an insight to star-formation activity through far-infrared radiation.
It also enables molecular hydrogen formation and traces the magnetic
field via emission and extinction polarization. The basic properties
of dust (optical, alignment etc.) strongly depend on its size distribution.
The latter evolves as the result of grain collisions, whose frequency
and consequences depend on grain relative velocities.

Grain-grain collisions can have various outcomes, e.g., coagulation,
cratering, shattering, and vaporization. For collisions with \( \delta v\leq 10^{-3} \)km/s,
grains are likely to stick or coagulate, as the potential energy due
to surface forces exceeds the initial center of mass kinetic energy.
Coagulation is considered the mechanism to produce large grains in
dark clouds and accretion disks. Collisions with \( \delta v\geq 20 \)km/s
have sufficient energy to vaporize at least the smaller of the colliding
grains (Draine 1985). Shattering threshold is much smaller than vaporization
one so that shattering dominates over vaporization in moderate energy
grain-grain collisions.

It is likely that some features of the grain distribution (e.g. Kim,
Martin \& Hendry 1994), e.g., the cutoff at large size, are the result
of fragmentation (Biermann \& Harwit 1980). Even low-velocity grain
collisions may have dramatic consequences by triggering grain mantle
explosion (Draine 1985).

Various processes can affect the velocities of dust grains. Radiation,
ambipolar diffusion, and gravitational sedimentation all can bring
about a dispersion in grain velocities. It is widely believed that,
except in special circumstances (e.g., near a luminous young star,
or in a shock wave), none of these processes can provide substantial
random velocities so as to affect the interstellar grain population
via collisions (Draine 1985). Nevertheless, it was speculated in de
Oliveira-Costa et al. (2000) that starlight radiation can produce
the segregation of different sized grains that is necessary to explain
a poor correlation of the microwave and \( 100\mu m \) signals of
the foreground emission (Mukherjee et al. 2001). If true it has big
implications for the CMB foreground studies. However, the efficiency
of this segregation depends on grain random velocities, which we study
in this paper.

Interstellar gas is turbulent (see Arons \& Max 1975, Scalo 1987,
Lazarian, Pogosyan \& Esquivel 2002). Turbulence was invoked by a
number of authors (see Kusaka et al. 1970, Volk et al. 1980, Draine
1985, Ossenkopf 1993, Weidenschilling \& Ruzmaikina 1994) to provide
substantial relative motions of dust particles. However, they discussed
hydrodynamic turbulence. It is clear that this picture cannot be applicable
to the magnetized ISM as the magnetic fields substantially affect
fluid dynamics. Moreover dust grains are charged, and their interactions
with magnetized turbulence is very different from the hydrodynamic
case. This unsatisfactory situation motivates us to revisit the problem
and calculate the grain relative motions in magnetized ISM. In what
follows, we use the model of MHD turbulence by Goldreich and Sridhar
(1995, henceforth GS95), which is supported by recent numerical simulations
(Cho \& Vishniac 2000, Maron \& Goldreich 2001, Cho, Lazarian \& Vishniac
2002a, henceforth CLV02). We apply our results to the cold neutral
medium (CNM) and a dark cloud to estimate the efficiency of coagulation,
shattering and segregation of grains.

\section{MHD Turbulence and Grain Motion}

Unlike hydrodynamic turbulence, MHD turbulence is anisotropic, with
eddies elongated along the magnetic field. This happens because it
is easier to mix the magnetic field lines perpendicular to their direction
rather than to bend them. The energy of eddies drops with the decrease
of eddy size (e.g. \( v_{l}\sim l^{1/3} \) for the Kolmogorov turbulence)
and it becomes more difficult for smaller eddies to bend the magnetic
field lines. Therefore the eddies get more and more anisotropic as
their sizes decrease. As eddies mix the magnetic field lines at the
rate \( k_{\bot }v_{k} \), where \( k_{\bot } \)is a wavenumber
measured in the direction perpendicular to the local magnetic field
and \( v_{k} \) is the mixing velocity at this scale, the magnetic
perturbations(waves) propagate along the magnetic field lines at the
rate \( k_{\parallel }V_{A} \) ,where \( k_{\parallel } \) is the
parallel wavenumber and \( V_{A} \) is the Alfv\'{e}n velocity.
The corner stone of the GS95 model is a critical balance between those
rates, i.e., \( k_{\bot }v_{k} \)\( \sim  \) \( k_{\parallel }V_{A} \),
which may be also viewed as coupling of eddies and wave-like motions.
Mixing motions perpendicular to the magnetic field lines are essentially
hydrodynamic (see CLV02) since magnetic field does not influence motions
that do not bend it. Therefore it is not surprising that the GS95
predicted the Kolmogorov one-dimensional energy spectrum in terms
of \( k_{\bot } \), i.e., \( E(k_{\bot })\sim k_{\bot }^{-5/3} \)(see
review by Cho, Lazarian \& Yan 2002, henceforth CLY02)\footnote{%
Two dimensional spectrum \( E(k_{\bot }, \)\( k_{\parallel }) \)
is presented in Cho, Lazarian \& Vishniac (2002a). 
}.

The GS95 model describes incompressible MHD turbulence. Maron \& Goldreich
(2001), Cho, Lazarian \& Vishniac (2002, henceforth CLV02) considered
pseudo Alfv\'{e}n mode, which is the limit of slow mode in incompressible
medium, and find a similar scaling relation. Recent research suggests
that the scaling is approximately true for the Alfv\'{e}nic and slow
mode in a compressible medium with Mach numbers(\( M\equiv V/C_{s} \))
of the order of unity (Lithwick \& Goldreich 2001, henceforth LG01,
Cho \& Lazarian 2002), which is also consistent with the analysis
of observational data (Lazarian \& Pogosyan 2000, Stanimirovic \&
Lazarian 2001, CLY02). In what follows we apply the GS95 scaling\footnote{%
Calculations in Cho \& Lazarian (2002) show that fast mode have spectrum
\( E(k)\sim k_{\bot }^{-3/2}, \) which is not very different from
the GS95 scaling. In the Appendix, we show that the motions that the
fast waves produced in low \( \beta =P_{gas}/P_{mag} \) plasma are
nearly perpendicular to the magnetic field. 
} to handle the problem of grain motions.

In MHD case, grain motions are different from those in hydrodynamic
turbulence and depend on the grain charge to mass ratio. If their
periods of Larmor motion \( \tau _{L} \) are longer than the gas
drag time \( t_{drag} \), the grains do not feel magnetic field directly.
Since the motions perpendicular to the local magnetic field are similar
to hydrodynamic motions, the estimates obtained in the Kolmogorov
hydrodynamic model should be essentially correct within a factor of
order unity. Otherwise, if the Larmor time \( \tau _{L} \) is shorter
than gas drag time \( t_{drag} \), grain perpendicular motions are
constrained by magnetic field.

Because of turbulence anisotropy, it is convenient to consider separately
grain motions parallel and perpendicular to the magnetic field. The
motions perpendicular to the magnetic field are influenced by Alfv\'{e}n
mode, while those parallel to the magnetic field are subjected to
the magnetosonic modes. The scaling relation for perpendicular motion
is \( v_{k}\propto k_{\perp }^{-1/3} \) (GS95). As the eddy turnover
time is \( \tau _{k}\propto (k_{\perp }v_{k})^{-1} \), the velocity
may be expressed as \( v_{k}\approx v_{max}\left( \tau _{k}/\tau _{max}\right) ^{1/2}, \)
where \( \tau _{max}=l_{max}/v_{max} \) is the time-scale for the
largest eddies, for which we adopt the fiducial values \( l_{max}=10 \)pc,
\( v_{max}=5 \)km/s.

If the Larmor time \( \tau _{L}=2\pi m_{gr}c/qB \) is shorter than
the gas drag time \( t_{drag} \), grain perpendicular motions are
constrained by the magnetic field. In this case, grains have a velocity
dispersion determined by the turbulence eddy whose turnover period
is \( \sim \tau _{L} \), while grains move with the magnetic field
on longer time scales. Since the turbulence velocity grows with the
eddy size, the largest velocity difference occurs on the largest scale
where grains are still decoupled. Thus, following the approach in
Draine (1985), we can estimate the characteristic grain velocity relative
to the fluid as the velocity of the eddy with a turnover time equal
to \( \tau _{L} \), \begin{equation}
\label{vperp}
v_{\perp }(a)={v^{3/2}_{max}\over l^{1/2}_{max}}(\rho _{gr})^{1/2}\left( {8\pi ^{2}c\over 3qB}\right) ^{1/2}a^{3/2},
\end{equation}
 and the relative velocity of grains to each other should be approximately
equal to the larger one of the grains' velocities, i.e., the larger
grain's velocity, \begin{eqnarray}
\delta v_{\perp }(a_{1},a_{2}) & = & {v^{3/2}_{max}\over l^{1/2}_{max}}(\rho _{gr})^{1/2}\left( {8\pi ^{2}c\over 3qB}\right) ^{1/2}[max(a_{1},a_{2})]^{3/2}\nonumber \\
 & = & 1.4\times 10^{5}cm/s(v_{5}a_{5})^{3/2}/(q_{e}l_{10}B_{\mu })^{1/2},\label{dvperp} 
\end{eqnarray}
 in which \( v_{5}=v_{max}/10^{5} \)cm/s, \( a_{5}=a/10^{-5} \)cm,
\( q_{e}=q/1 \)electron, \( l_{10}=l_{max}/10 \)pc, \( B_{\mu }=B/1\mu  \)G,
and the grain density is assumed to be \( \rho _{gr}=2.6 \)g/cm\( ^{-3} \)
.

Grain motions parallel to the magnetic field are induced by the compressive
component of slow mode with \( v_{\parallel }\propto k_{\parallel }^{-1/2} \)
(CLV02, LG01, CLY02). The eddy turnover time is \( \tau _{k}\propto (v_{\parallel }k_{\parallel })^{-1} \),
so the parallel velocity can be described as \( v_{\parallel }\approx v_{max}\tau _{k}/\tau _{max} \)\footnote{%
We assume that turbulence is driven isotropically at the scale \( l_{max} \). 
}. For grain motions parallel to the magnetic field the Larmor precession
is unimportant and the gas-grain coupling takes place on the translational
drag time \( t_{drag} \). The drag time due to collisions with atoms
is essentially the time for collision with the mass of gas equal to
the mass of grain, \( t^{0}_{drag}=(a\rho _{gr}/n)(\pi /8\mu kT)^{1/2} \),
where \( \mu  \) is the mass of gas species. The ion-grain cross-section
due to long-range Coulomb force is larger than the atom-grain cross-section
. Therefore, in the presence of collisions with ions, the effective
drag time decreases by the factor

\[
\alpha =\left[ \frac{1}{2}\sum _{i}\left( \frac{m_{i}}{m_{\textrm{H}}}\right) ^{1/2}\sum _{Z}f_{z}\left( \frac{Ze^{2}}{akT}\right) ^{2}\ln \left[ \frac{3(kT)^{3/2}}{2e^{3}|Z|(\pi xn_{\textrm{H}})^{1/2}}\right] \right] ^{-1}\]
 where \( x_{i} \) is the ionization of ion \( i \) with mass \( m_{i} \),
\( f_{Z} \) is the probability of the grain in the state with charge
\( Z \) (Draine \& Salpeter 1979). The characteristic velocity of
grain motions along the magnetic field is approximately equal to the
parallel turbulent velocity of eddies with turnover time equal to
\( t_{drag}=t_{drag}^{0}/\alpha  \)

\begin{equation}
\label{vpara}
v_{\parallel }(a)=\alpha ^{-1}{v^{2}_{max}\over l_{max}}\left( \frac{\rho _{gr}}{4n}\right) \left( {2\pi \over \mu kT}\right) ^{1/2}a,
\end{equation}
 and the relative velocity of grains for \( T_{100}=T/100 \)K is

\begin{eqnarray}
\delta v_{\parallel }(a_{1},a_{2}) & = & \alpha ^{-1}{v^{2}_{max}\over l_{max}}\left( \frac{\rho _{gr}}{4n}\right) ({2\pi \over \mu kT})^{1/2}[max(a_{1},a_{2})]\nonumber \\
 & = & (1.0\times 10^{6}cm/s)\alpha v_{5}^{2}a_{5}/(nl_{10}T_{100}^{1/2}),
\end{eqnarray}

When \( \tau _{L}>t_{drag} \), grains are no longer tied to the magnetic
field. Since at a given scale, the largest velocity dispersion is
perpendicular to the magnetic field direction, the velocity gradient
over the grain mean free path is maximal in the direction perpendicular
to the magnetic field direction. The corresponding scaling is analogous
to the hydrodynamic case, which was discussed in Draine (1985): \begin{eqnarray}
\delta v(a_{1},a_{2}) & = & {v^{3/2}_{max}\over l^{1/2}_{max}}t_{drag}^{1/2}\nonumber \\
 & = & \alpha ^{-1/2}{v^{3/2}_{max}\over l^{1/2}_{max}}\left( \frac{\rho _{gr}}{4n}\right) ^{1/2}\left( {2\pi \over \mu kT}\right) ^{1/4}[max(a_{1},a_{2})]^{1/2}.\label{HD} 
\end{eqnarray}

An important difference between hydrodynamic and MHD turbulence is
the presence of the ambipolar damping of turbulence in MHD case.(see
Cho, Lazarian \& Vishniac 2002b). If the mean free path for a neutral
particle, \( l_{n} \), in a partially ionized gas with density \( n_{tot}=n_{n}+n_{i} \)
is much less than the size of the eddies in consideration, i.e. \( l_{n}k_{\bot }\ll 1 \),
the damping time is

\begin{equation}
\label{tdamp}
t_{damp}\sim \nu _{n}^{-1}k_{\perp }^{-2}\sim \left( n_{tot}/n_{n}\right) \left( l_{n}v_{n}\right) ^{-1}k_{\perp }^{-2},
\end{equation}
 where \( \nu _{n} \) is effective viscosity produced by neutrals\footnote{%
The viscosity due to ion-ion collisions is typically small as ion
motions are constrained by the magnetic field. 
}. The mean free path of a neutral \( l_{n} \) is influenced both
by collisions with neutrals and with ions. The rate of a neutral colliding
with ions is proportional to the density of ions, while the rate of
a neutral colliding with other neutrals is proportional to the density
of neutrals. The drag coefficient for neutral-neutral collisions is
\( \sim 1.7\times10 ^{-10}T \)(K)\( ^{0.3} \) cm\( ^{3} \) s\( ^{-1} \)
(Spitzer 1978), while for neutral-ion collisions it is \( \sim {\langle v_{r}\sigma _{in}\rangle }\approx 1.9\times 10^{-9} \)
cm\( ^{3} \) s\( ^{-1} \) (Draine, Roberge \& Dalgarno 1983). Thus
collisions with other neutrals dominate for \( n_{i}/n_{n} \) less
than \( \sim 0.09T^{0.3} \). In the present paper we consider cold
gas and therefore the influence of ions on \( l_{n} \) is disregarded.

Turbulent motions cascade down till the eddy turnover is of the order
of \( t_{damp} \). Thus the turbulence cutoff in neutral medium is
\begin{equation}
\label{cutoff}
\tau _{c}\simeq \left( \frac{l_{n}}{v_{n}}\right) \left( \frac{v_{n}}{v_{max}}\right) ^{\frac{3}{2}}\left( \frac{l_{max}}{l_{n}}\right) ^{\frac{1}{2}}\left( \frac{V_{A}}{v_{max}}\right) ^{\frac{1}{2}}\left( \frac{n_{n}}{n_{tot}}\right) ,
\end{equation}
 where \( v_{n} \) and \( V_{A} \) are, respectively, the velocity
of a neutral and Alfv\'{e}n velocity. It is easy to see that for
\( \tau _{c} \) longer than either \( t_{drag} \) or \( \tau _{L} \)
the grain motions get modified. A grain samples only a part of the
eddy before gaining the velocity of the ambient gas. In GS95 picture,
the shear rate \( dv/dl \) increases with the decrease of eddy size.
Thus for \( \tau _{c}>max\{t_{drag},\tau _{L}\} \), these smallest
available eddies are the most important for grain acceleration. Consider
first the perpendicular motions. If \( v_{c} \) is the velocity of
the critically damped eddy, the distance traveled by the grain is
\( \bigtriangleup l\sim v_{c}\times min\lbrace t_{drag},\tau _{L}\rbrace  \).
The shear rate \( dv/dl \) perpendicular to the magnetic field is
\( \tau _{k}^{-1} \). Thus the grain experiences the velocity difference
in the direction perpendicular to the magnetic field

\begin{equation}
\label{vperp'}
v_{\perp }\sim \bigtriangleup l\times \frac{dv}{dl}\sim \frac{v_{c}}{\tau _{c}}\times min\lbrace t_{drag},\tau _{L}\rbrace .
\end{equation}
 For the parallel motions, \( \bigtriangleup l\sim v_{c}\times t_{drag} \).
And if noticing the critical balance in GS95 model \( k_{\parallel }V_{A}\sim k_{\perp }v_{\perp }=\tau _{k}^{-1} \),
the largest shear rate along the magnetic field should be \( dv/dl=v_{c}k_{\parallel }\sim v_{c}/(V_{A}\tau _{c}) \).
Therefore, in the parallel diretion, grain experiences a velocity
difference \( V_{A}/v_{c} \) times smaller, i.e.,

\[
v_{\parallel }\sim \frac{v_{c}^{2}}{V_{A}}\times \frac{t_{drag}}{\tau _{c}}.\]

\section{Discussion }

\subsection{Shattering and Coagulation }

Consider the cold neutral medium (CNM) with temperature \( T=100 \)K,
density \( n_{\textrm{H}}=30 \)cm\( ^{-3} \), electron density \( n_{e}=0.045 \)cm\( ^{-3} \),
magnetic field \( B\sim 1.3\times 10^{-5} \)G (Weingartner \& Draine
2001a, hereafter WD01a). To account for the Coulomb drag, we use the
results by WD01a and get the modified drag time \( t_{drag}=\alpha ^{-1}t^{0}_{drag} \).
Using the electric potentials in Weingartner \& Draine (2001b), we
get grain charge and \( \tau _{L} \).

For the parameters given above, we find that \( t_{drag} \) is larger
than \( \tau _{c} \) for grains larger than \( 10^{-6} \)cm, \( \tau _{L} \)
is smaller than \( \tau _{c} \) even for grains as large as \( 10^{-5} \)cm.
Here, we only consider grains larger than \( 10^{-6} \)cm, which
carry most grain mass (\( \sim 80\% \)) in ISM, so we can still use
Eq.(\ref{vpara}) to calculate grain parallel velocities and Eq.(\ref{vperp})
to get the perpendicular velocity for grain larger than \( 10^{-5} \)cm.
Nevertheless, the perpendicular velocities of grains smaller than
\( 10^{-5} \)cm should be estimated according to Eq.(\ref{vperp'}),

\begin{equation}
\label{vperpa'}
v'_{\perp }(a)=v_{max}\left( \frac{\tau _{c}}{\tau _{max}}\right) ^{1/2}\left( \frac{\tau _{L}}{\tau _{c}}\right) =v_{\perp }(a)\left( \frac{\tau _{L}}{\tau _{c}}\right) ^{1/2},
\end{equation}
 as where \( v_{\perp }(a) \) is given by Eq.(\ref{vperp}). The
results are shown in Fig.1. The relative velocity, as we discussed
earlier, is dependent on the velocity of the larger grain. Thus if
\( max\lbrace a_{1},a_{2}\rbrace  \) is larger than \( 10^{-5} \)cm,
the grain relative velocity should be given by Eq.(\ref{dvperp}).
Otherwise, we should use Eq.(\ref{vperpa'}) to estimate the grain
relative velocity.

The critical sticking velocity were calculated in Chokshi et al. (1993)(see
also Dominik \& Tielens 1997).\footnote{%
There are obvious misprints in the numerical coefficient of Eq.(7)
in Chokshi et al.(1993) and the power index of Young's modulus in
Eq.(28) of Dominik \& Tielens (1997). 
} However, experimental work by Blum (2000) shows that the critical
velocity is an order of magnitude larger than the theoretical calculation.
Thus the collisions can result in coagulation for small silicate grains
(\( \leq 3\times 10^{-6} \)cm).

According to our result, grains won't be shattered if the shattering
thresholds for silicate is \( 2.7 \)km/s (Jones et al. 1996). Nevertheless,
the grain velocities are strongly dependent on the maximal velocity
of tubulence \( v_{max} \) at the injection scale, which is highly
uncertain. The value \( v_{max}=5 \)km/s we use here is a conservative
estimate. We expect grains would acquire larger velocities, therefore,
more likely to be shattered during collisions if the \( v_{max} \)
is larger. For instance, we will get a cutoff \( 6\times 10^{-5} \)cm
due to shattering if \( v_{max}=10 \)km/s.

For a dark cloud, the situation is different. As the density increases,
the drag by gas becomes stronger. Consider a typical dark cloud with
temperature \( T=20 \)K, density \( n_{\textrm{H}}=10^{4} \)cm\( ^{-3} \)
(Chokshi et al. 1993) and magnetic field \( B\sim 2.3\times 10^{-4} \)G.
Assuming that dark clouds are shielded from radiation, grains get
charged by collisions with electrons: \( <q>=0.3(r/10^{-5} \)cm)
electrons. The ionization in the cloud is \( \chi =n_{e}/n_{tot}\sim 10^{-6} \)
and the drag by neutral atoms is dominant. From Eq.(\ref{cutoff})
and the expression for the drag time and the Larmor time, we find
\( \tau _{L}<t_{drag} \) for grains of sizes between \( 10^{-6} \)cm
and \( 4\times 10^{-6} \)cm, and \( t_{drag}<\tau _{L} \) for grains
larger than \( 4\times 10^{-6} \)cm. In both cases, turbulence cutoff
\( \tau _{c} \) is smaller than \( t_{drag} \) and \( \tau _{L} \).
Thus for the smaller grains, we use Eq.(\ref{vperp}),(\ref{vpara})
to estimate grain velocities. For larger grains, the fact \( t_{drag}<\tau _{L} \)
means that grain velocities are similiar to the hydrodynamic case
and Eq(\ref{HD}) is used (Fig 1b).

Our results for dark clouds show only a slight difference from the
earlier hydrodynamic estimates. Since the drag time \( t_{drag}\propto n^{-1} \),
Larmor time \( \tau _{L}\propto B^{-1}\propto n^{-1/2} \), the grain
motions get less affected by the magnetic field as the cloud becomes
denser. Thus we agree with Chokshi's et al. (1993) conclusion that
densities well in excess of \( 10^{4} \)cm\( ^{-3} \) are required
for coagulation to occur. Shattering will not happen because the velocities
are small, so there are more large grains in dark clouds. This agrees
with observations (see Mathis 1990).

Supersonic grain motion in respect to gas may result in grain alignment
(see Lazarian 2000 for a review). Our present results testify that
only grains with size larger than \( \sim 10^{-4} \)cm experience
such velocities for the typical parameters of the CNM. Such grains
are marginally important for starlight polarization (Kim \& Martin
1995). This, however, does not preclude grain alignment via streaming
in particular circumstances, e.g. in shocks and outflows (Roberge
\& Hanany 1990, Lazarian 1994) For low \( \beta (\beta =P_{gas}/P_{mag}) \)
is the ratio of gas pressure and magnetic pressure) media, the supresonic
motions would correspond to fast mode and Alfv\'{e}nic mode (Alfv\'{e}n
\& F\( \ddot{a} \)lthmmar 1963). Alfv\'{e}nic mode have velocity
perpendicular to \textbf{B} and fast mode have velocity nearly perpendicular
to \textbf{B}. As the mechnical alignment\footnote{%
For grains rotating thermally the Gold alignment theory (Gold 1952,
Roberge et al. 1993, Lazarian1997) is applicable, while cross sectional
and crossover (Lazarian 1995, Lazarian \& Efroimsky 1996) is applicable. 
} minimizes grain cross section to the flow, the alignment would happen
with the long axis perpendicular to the magnetic field.

\subsection{Grain Segregation and Turbulent Mixing}

Our results are also relevant to grain segregation. Grains are the
major carrier of heavy elements in the ISM. The issue of grain segregation
may have significant influence on the ISM metallicity. Subjected to
external forcing, e.g., due to radiation pressure, grains gain size-dependent
velocities with respect to gas. WD01a have considered the forces on
dust grains exposed to anisotropic interstellar radiation fields.
They included photoelectric emission, photodesorption as well as radiation
pressure, and calculated the drift velocity for grains of different
sizes. The velocities they got for silicate grains in the CNM range
from \( 0.1 \)cm/s to \( 10^{3} \)cm/s. Fig.1 shows that the turbulence
produces larger velocity dispersions.\footnote{%
If reconnection is fast (see Lazarian \& Vishniac 1999), the mixing
of grains over large scales is provided by turbulent diffusivity\( \sim v_{max}l_{max} \).
On small scales the grain decoupled motions are important. 
} Thus the grain segregation of very small and large grains speculated
in de Oliveira-Costa et al. (2000) is unlikely to happen for typical
CNM conditions.

A different mechanism of driving grain motions is a residual imbalance
in {}``rocket thrust{}'' between the opposite surfaces of a rotating
grain (Purcell 1979). This mechanism can provide grain relative motions
and preferentially move grains into molecular clouds. It is easy to
see that due to averaging caused by grain rotation, the rocket thrust
is parallel to the rotation axis. Three causes for the thrust were
suggested by Purcell (1979): spatial variation of the accommodation
coefficient for impinging atoms, photoelectric emission, and H\( _{2} \)
formation. The latter was shown to be the strongest among the three.
The uncompensated force in this case arises from the difference of
the number of catalytic active sites for H\( _{2} \) formation on
the opposite grain surfaces. The nascent H\( _{2} \) molecules leave
the active sites with kinetic energy \( E \) and the grain experiences
a push in the opposite directions. The number of active sites varies
from one grain to another, and we should deal with the expectation
value of the force for a given distribution of active sites.

Adopting the approach in Lazarian \& Draine (1997), we get the mean
square root force of H\( _{2} \) formation on a grain in the shape
of a square prism with dimensions \( b\times b\times a \) (with \( b>a \)
)

\begin{equation}
\label{H}
\langle F_{z{\textrm{H}}}\rangle =r^{3/2}(r+1)^{1/2}\gamma (1-y)n_{\textrm{H}}v_{\textrm{H}}a^{2}\left( \frac{2m_{\textrm{H}}E}{\nu }\right) ^{1/2},
\end{equation}
 where \( r=b/2a, \), \( n_{\textrm{H}}\equiv n({\textrm{H}})+2n({\textrm{H}}_{2}) \),
\( y=2n({\textrm{H}}_{2})/n_{\textrm{H}} \) is the \( {\textrm{H}_{2}} \)
fraction, \( \gamma  \) is the fraction of impinging H atoms and
\( \nu  \) is the number of active sites over the grain surface.
Due to internal relaxation of energy (Purcell 1979, Spitzer \& McGlynn
1979, Lazarian \& Efroimsky 1999, Lazarian \& Draine 1999a,b) the
grain rotational axis tends\footnote{%
This statement is true up to grain thermal wiggling (Lazarian 1994,
Lazarian \& Roberge 1997) and occasional flipping (Lazarian \& Draine
1999). 
} to be perpendicular to the largest \( b-b \) surface, the uncompensated
force above is parallel to the rotational axis. The other components
of force are being averaged out due to grain fast rotation. The expected
grain velocity is \( v=\langle F_{zH}\rangle t_{drag}/m. \) In the
CNM we consider, \( y=0 \), adopting the characteristic values in
Lazarian \& Draine (1997), \( r=1, \) \( \gamma =0.2 \), \( E=0.2 \)eV,
and the density of active sites \( 10^{11} \)cm\( ^{-2} \) so that
\( \nu =80(a/10^{-5} \)cm\( )^{2}r(r+1) \),\footnote{%
The number of H\( _{2} \) formation sites is highly uncertain. It
may also depend on the interplay of the processes of photodesorption
and poisoning (Lazarian 1995, 1996). 
} we get the {}``optimistic{}'' velocity, \( v\simeq 330(10^{-5} \)cm\( /\sqrt{a}) \)cm/s,
shown in Fig 1. The scaling is approximate due to the complexity of
coefficient \( \alpha  \) (see WD01a Fig.16). For maximal active
site density \( 10^{15} \)cm\( ^{-2} \), we get the lower boundary
of grain velocity \( v\simeq 3.3(10^{-5} \)cm\( /a)^{1/2} \)cm/s.
Such velocities may have noticable effects. For instance, H\( _{2} \)
forces and therefore H\( _{2} \) driven velocities are expected to
drop in molecular clouds, resulting in preferential diffusion of heavy
elements into molecular clouds. Due to the size dependence of the
velocities, one could also expect a segregation of grains of different
sizes.

For small grains (\( <3\times 10^{-6} \)cm), the {}``optimistic{}''
velocity obtained from H\( _{2} \) formation is higher than that
arising from our turbulent estimates. Does this mean that such grains
will be segregated? Our calculation implicitly assumes that grains
do not flip. Therefore, they experience constant forcing through \( t_{drag} \).
However, Lazarian \& Draine (1999a) have shown that subjected to H\( _{2} \)
torques alone small grains \( \leq 10^{-5} \)cm should experience
frequent thermal flipping, which means that the forces are averaged
out by this effect. This flipping results from coupling of grain rotational
and vibrational degrees of freedom through internal relaxation. The
Barnett relaxation discovered by Purcell (1979) was used for the calculations.
Further research by Lazarian \& Draine (1999b) resulted in a discovery
of nuclear relaxation that made flipping efficient for grains up to
\( 10^{-4} \)cm. Does this mean that the H\( _{2} \) forces on interstellar
grains are completely averaged out? Probably not. The flipping efficiency
depends on the value of the grain angular momentum (Lazarian \& Draine
1999a). If grains are already spun up to sufficient velocities, they
get immune to thermal flipping. Radiative torques (Dolginov \& Mytrophanov
1975, Draine 1996, Draine \& Weingartner 1996) provide a possibility
of spinning grains up to high velocities. For them to be active, the
grain size should be comparable to the wavelength. Therefore, for
a typical interstellar diffuse radiation field radiative torques are
expected to spin up grains with sizes larger than \( \sim 4\times 10^{-6} \)cm.
They will also align grains with rotational axes parallel to the magnetic
field. Thus grains should acquire velocities along the magnetic field
lines and the corresponding velocities should be compared with those
arising from turbulent motions parallel to the magnetic field. It
is clear from Fig.1 that for the chosen set of parameters the effect
of H\( _{2} \) thrust is limited. All in all, we conclude that the
radiation effects and H\( _{2} \) thrust are not efficient for segregating
grains in typical ISM conditions.

\begin{figure}
\plottwo{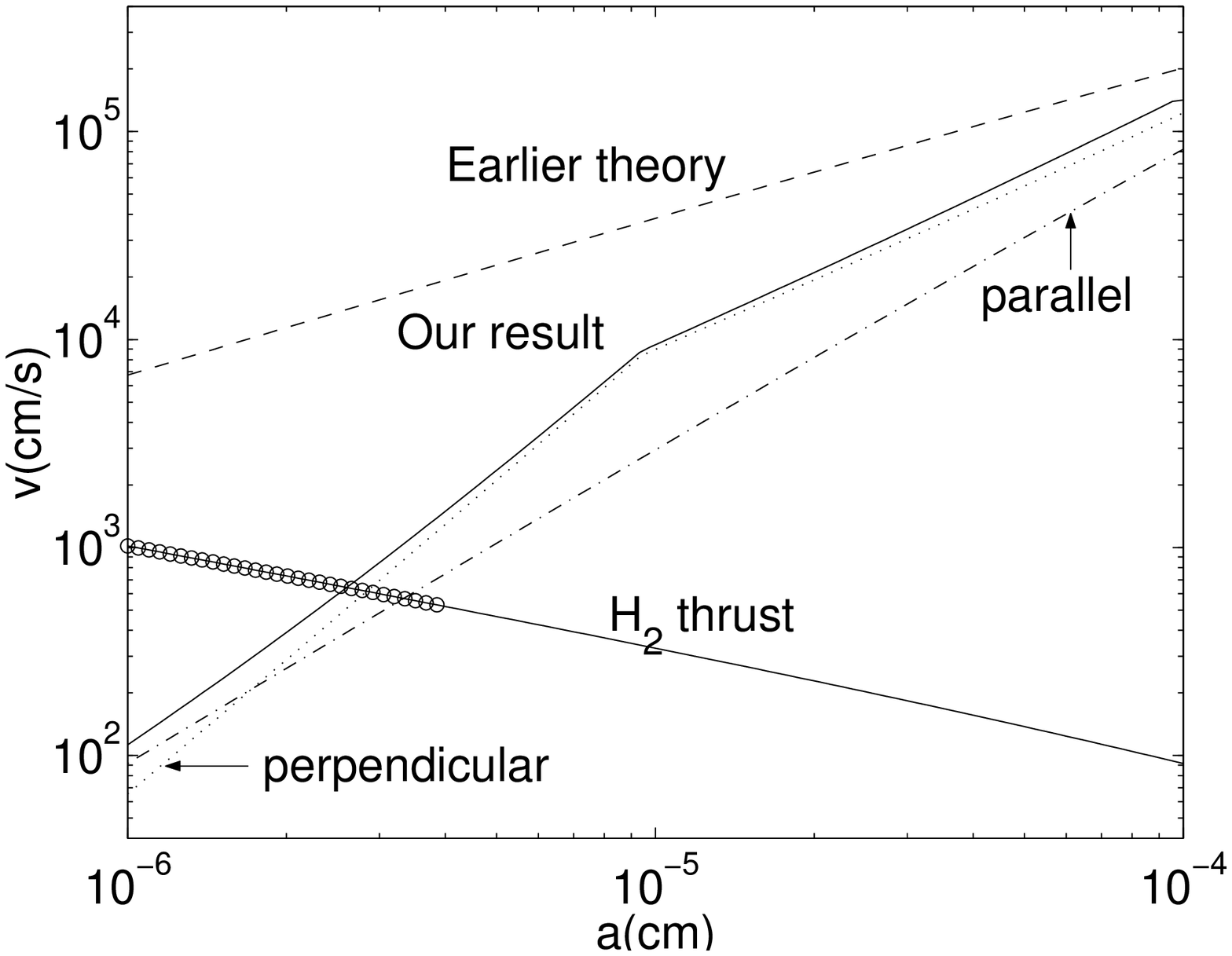}{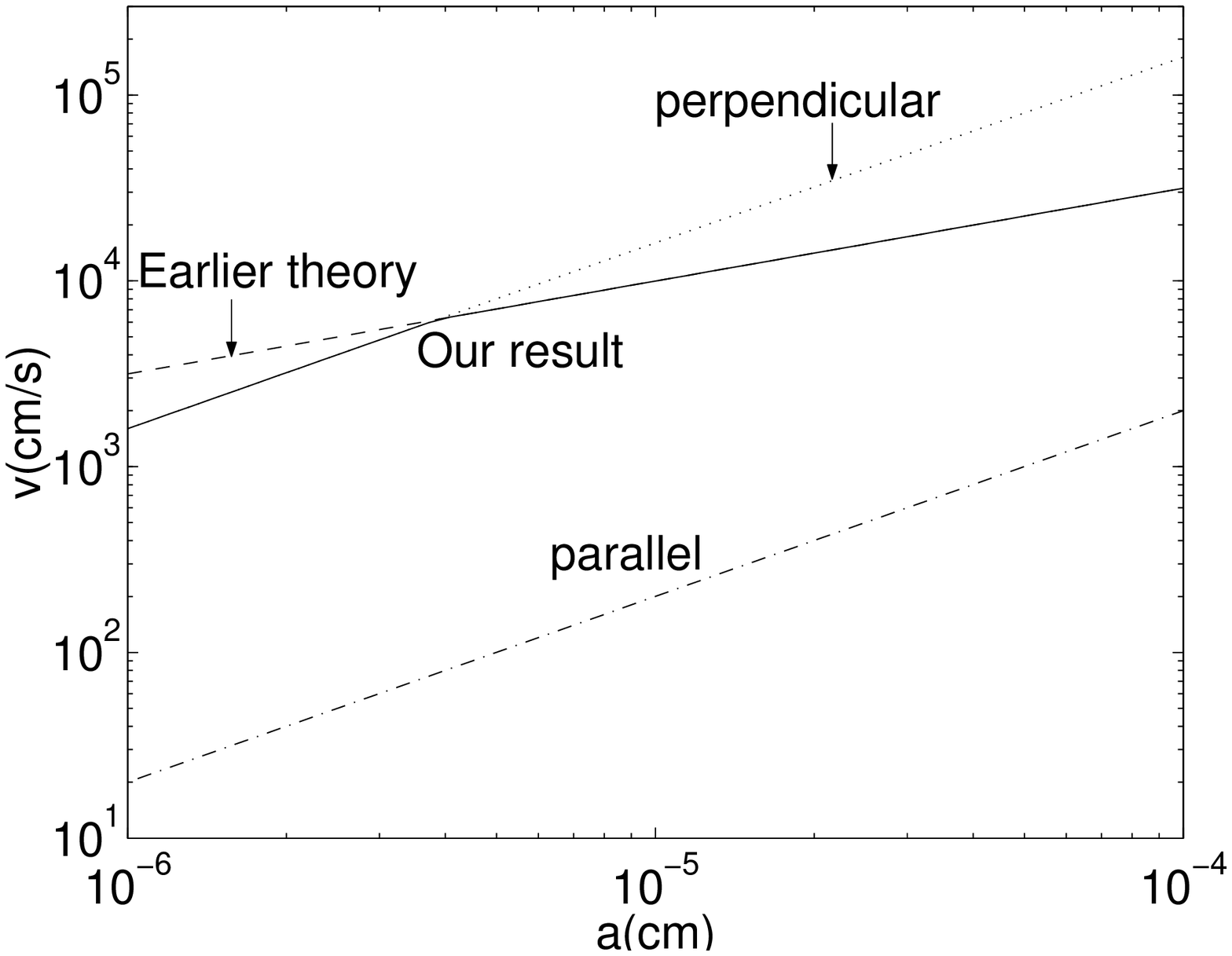}
\caption{Grain relative velocities as a function of radii (solid line) (a)
in CNM, (b) in DC. Dashdot line represents parallel velocity due to
the drag by compressible modes, dotted line refers to perpendicular
velocity from the contribution of the drag by Alfv\'{e}n mode, also
plotted is the earlier hydrodynamic result (dashed line). The change
of the slope in (a) is due to the cutoff of turbulence by ambipolar
diffusion. In (b), for grains larger than \protect\protect\( \sim 4\times 10^{-6}\protect \protect \)cm,
the grains velocities are similiar with the hyro case (see text).
The grain velocity driven by H\protect\protect\protect\( _{2}\protect \protect \protect \)
thrust is plotted to illustrate the issue of grain segregation in
the CNM (see text), the part marked by 'o' is nonphysical because
thermal flipping is not taken into account. }
\end{figure}

It is possible to show that for the CNM the effect of variation of
accommodation coefficient is less important than the H\( _{2} \)
formation. Photoelectric emission does not vanish as grains flip and
the corresponding motion was considered by WD01a, but the velocity
due to the effect is smaller than that arising from turbulent motions.
Thus, we think segregation is marginal in such a CNM. Some effect
on metallicity might be present and worth quantifying elsewhere.

The percentage of atomic hydrogen is reduced in dark clouds, and the
radiation field is weak. Thus the only force we should pay attention
to is the bombardment by gas atoms. According to Purcell (1979), the
force should be \( \langle F_{zb}\rangle =\sqrt{8a^{2}/s}\bigtriangleup T/2\delta s^{2}, \)
for a grain with a deviation \( \delta  \) in accommodation coefficient
by a scale of \( s \) if the temperature difference between gas and
grain is \( \bigtriangleup T \). This temperature difference is usually
small (\( <10 \) K). Moreover, because of insufficient radiation
grains are expected to flip frequently. Therefore, the velocities
driven by the variation of the accommodation coefficient are always
much smaller than those due to turbulence drag so that grains in dark
clouds should be fully mixed.

\subsection{Resonant Grain Acceleration }

In the treatment above we disregarded the possibility of direct acceleration
of grains through their interactions with the fluctuating electromagnetic
field in the MHD turbulence. As we mentioned in \( \S  \)2, the turbulent
motions parallel to the magnetic field are like MHD waves. The charged
grains interact with the waves effectively when the wave frequency
is a multiple of the gyrofrequency of the grain\footnote{%
We should take into account resonance broadening if the dynamical
time scale of turbulence is comparable to the time scale of grain
motion. 
}, \( \omega -k_{\parallel }v\mu =n\Omega  \), this is so called gyroresonance.
This resonant process is important for a highly ionized medium. In
our next paper we will discuss the stochastic grain accleration by
gyroresonance with different MHD modes in warm ionized medium. In
partially ionized medium, the turbulence is damped due to ion-neutral
collisions (see the discussion around Eq.(\ref{tdamp})). For the
CNM we consider, we find in that the turbulence is cut off at \( \tau _{c}=2\pi /\omega  \)
beyond the Larmor time \( \tau _{L}=2\pi /\Omega  \) of the grain
less than \( 10^{-5} \)cm (see Fig. 1a). Since the grain(\( <10^{-5} \)cm)
velocities are much smaller than Alfv\'{e}n velocity \( V_{A}=10^{5} \)cm/s,
the resonant condition is not satisfied for these grains. Indeed,
there are still some magnetic structures below the cutoff according
to Cho, Lazarian \& Vishniac (2002b). However, these structure are
static so that they do not contribute to grain acceleration. Therefore,
while the velocities of larger grains(\( >10^{-5} \)cm) may be modified
by gyroresonance, the velocities of those smaller grains(\( <10^{-5} \)cm)
should remain the same as shown in Fig.1.

\section{Summary}

We have calculated relative motions of dust grains in a magnetized
turbulent fluid taking into account turbulence anisotropy, turbulence
damping and grain coupling with the magnetic field. We find that these
effects decrease the relative velocities of dust grains compared to
the earlier hydrodynamic-based calculations. The difference is substantial
in CNM, but less important for dark clouds. For CNM we find that coagulations
of silicate grains happen for sizes \( \leq 3\times 10^{-6} \)cm.
The force due to H\( _{2} \) formation on grain surface might drive
small grains (\( <3\times 10^{-6} \)cm) to larger velocities if thermal
flipping of grains is suppressed by grain rapid rotation. This may
be possible in the presence of radiative torques. However, radiative
torques are suppressed for grains smaller than wavelength and therefore
grains are expected to be well mixed due to turbulence.

We are grateful to John Mathis for reading the manuscript and many
important comments. The research is supported by the NSF grant AST0125544.

\appendix
\section{Angle between \( \textbf {B} \) and \( {\textbf {v}} \)}

A fundamental question arises from the fact, that in MHD turbulence
wave vectors are not aligned along magnetic field lines, as this is
the case for pure Alfv\'{e}nic waves. We need to analyze relative
position of three vectors: magnetic field vector \( \textbf {B} \),
wave vector \( \textbf {k} \), and the displacement velocity vector
\( \textbf {v} \). In what following, we shall study how the angle
\( \gamma  \) between \( {\textbf {v}} \) and \( \textbf {B} \)
changes with plasma \( \beta =P_{gas}/P_{mag}=2C_{s}^{2}/V_{A}^{2} \)
.

\begin{figure}
\centering \leavevmode
\resizebox*{0.33\columnwidth}{!}{\includegraphics{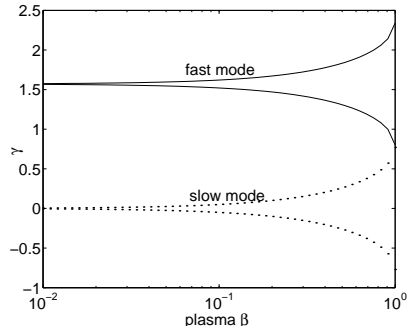} }

\caption{The range of angles between \protect\protect\( \textbf {B}\protect \protect \)
and \protect\protect\( {\textbf {v}}\protect \protect \) The solid
line refers to the fast mode, the dotted line is plotted for fast
mode. This range is produced when angle between \protect\protect\( {\textbf {k}}\protect \protect \)
and \protect\protect\( {\textbf {B}}\protect \protect \) changes
in the range \protect\protect\( 0\protect \protect \) to \protect\protect\( \pi \protect \protect \).\label{fm2}}
\end{figure}

It is shown in Alfv\'{e}n H. \& F\"{a}lthmmar (1963) that the angle
\( \Psi  \) between \( {\textbf {v}} \) and \( \textbf {k} \) can
be expressed as follows: \begin{equation}
\label{al1}
\tan \Psi =\frac{\sin \alpha \cos \alpha }{\cos ^{2}\alpha -v_{p}^{2}/V_{A}^{2}},
\end{equation}
 where \( v_{p} \) is the phase velocity, which is related to the
Alfv\'{e}nic velocity \( V_{A} \) and the sound velocity \( C_{s} \)
through the dispersion relation \begin{equation}
\label{al2}
v_{p}^{4}-(V_{A}^{2}+C_{s}^{2})v_{p}^{2}+C_{s}^{2}V_{A}^{2}\cos ^{2}\alpha =0.
\end{equation}
 Solving this equation in respect to \( \epsilon =v_{f}^{2}/v_{A}^{2} \),
\begin{equation}
\label{al3}
\epsilon (\beta )=\frac{1}{2}\left( 1+\beta /2\pm \sqrt{(1-\beta /2)^{2}+2\beta \sin ^{2}\alpha }\right) ,
\end{equation}
 where '\( + \)' gives the result for fast mode and '\( - \)' represents
slow mode. Thus the angle \( \gamma  \) can be calculated as \begin{equation}
\label{al5}
\gamma =\alpha -\arctan \frac{\sin \alpha \cos \alpha }{\cos ^{2}\alpha -\epsilon (\xi )}
\end{equation}
 and the corresponding plot is shown in Fig.~\ref{fm2}. It is evident
that for low \( \beta  \) plasma the velocity \( v \) of the fast
mode is directed nearly perpendicular to \( \textbf {B} \) whatever
is the direction of \( \textbf {k} \), while the velocity of the
slow mode is nearly parallel to the magnetic field. So the parallel
motions we got from slow mode are essentially correct, while the perpendicular
motions are also subjected to fast mode. Since the scaling of fast
mode according to Cho \& Lazarian (2002) is \( v_{k}\propto k^{-1/4}, \)
which is not so different from the scaling we use, \( v_{k}\propto k^{-1/3} \),
we expect that our result for perpendicular motions wouldn't be affected
substantially.

\end{document}